\documentclass[10pt,letterpaper]{article}

\usepackage{cogsci}

\cogscifinalcopy
\usepackage[flushleft]{threeparttable}
\usepackage{amsmath}
\usepackage{tipa}

\usepackage{pslatex}
\usepackage{graphicx}
\usepackage{xcolor}
\usepackage{apacite}
\usepackage{float} 
\usepackage{amsfonts}

\title{Characterizing the Interaction of Cultural Evolution Mechanisms in\\ Experimental Social Networks}

\author{
{\large \bf Raja Marjieh$\textsuperscript{1,*}$, Manuel Anglada-Tort$\textsuperscript{2,*}$, Thomas L. Griffiths$\textsuperscript{1,3}$, Nori Jacoby$\textsuperscript{4}$} \\
  $\textsuperscript{1}$Department of Psychology, Princeton University\\
  $\textsuperscript{2}$Department of Psychology, Goldsmiths College, University of London\\
  $\textsuperscript{3}$Department of Computer Science, Princeton University\\
  $\textsuperscript{4}$Department of Psychology, Cornell University\\
  $\texttt{raja.marjieh@princeton.edu; m.angladatort@gold.ac.uk;}$ \\ $\texttt{tomg@princeton.edu; kj338@cornell.edu}$ \\
  $\textsuperscript{*}$Equal contribution. 
  } 

\begin{document}
\maketitle
\begin{abstract}
Understanding how cognitive and social mechanisms shape the evolution of complex artifacts such as songs is central to cultural evolution research. Social network topology (what artifacts are available?), selection (which are chosen?), and reproduction (how are they copied?) have all been proposed as key influencing factors. However, prior research has rarely studied them together due to methodological challenges. We address this gap through a controlled naturalistic paradigm whereby participants ($N=2404$) are placed in networks and are asked to iteratively choose and sing back melodies from their neighbors. We show that this setting yields melodies that are more complex and more pleasant than those found in the more-studied linear transmission setting, and exhibits robust differences across topologies. Crucially, these differences are diminished when selection or reproduction bias are eliminated, suggesting an interaction between mechanisms. These findings shed light on the interplay of mechanisms underlying the evolution of cultural artifacts.

\end{abstract}

\section{Introduction}
Complex cultural systems -- such as technology \cite{derex2015foundations, thompson2022complex}, language \cite{kirby2008cumulative, raviv2019larger}, dance \cite{laland2016evolution}, and music \cite{anglada2023large, ravignani2016musical} -- emerge from the interaction of multiple mechanisms in social environments. For example, a new dance move may gain popularity when a famous person promotes it on social media, but this popularity may fade away if the dance move proves too difficult to learn or if it is too awkward to imitate in public. 
Understanding the mechanisms that shape the global properties of complex cultural artifacts has been central to numerous studies in anthropology \cite{henrich2016secret}, psychology \cite{tomasello2009cultural}, biology \cite{whiten2019cultural}, sociology \cite{salganik2006experimental}, linguistics \cite{gray2003language, kirby2008cumulative}, and cultural evolution research \cite{mesoudi2011cultural}.
\begin{figure}[htp]
\begin{center}
\includegraphics[width=0.85\linewidth]{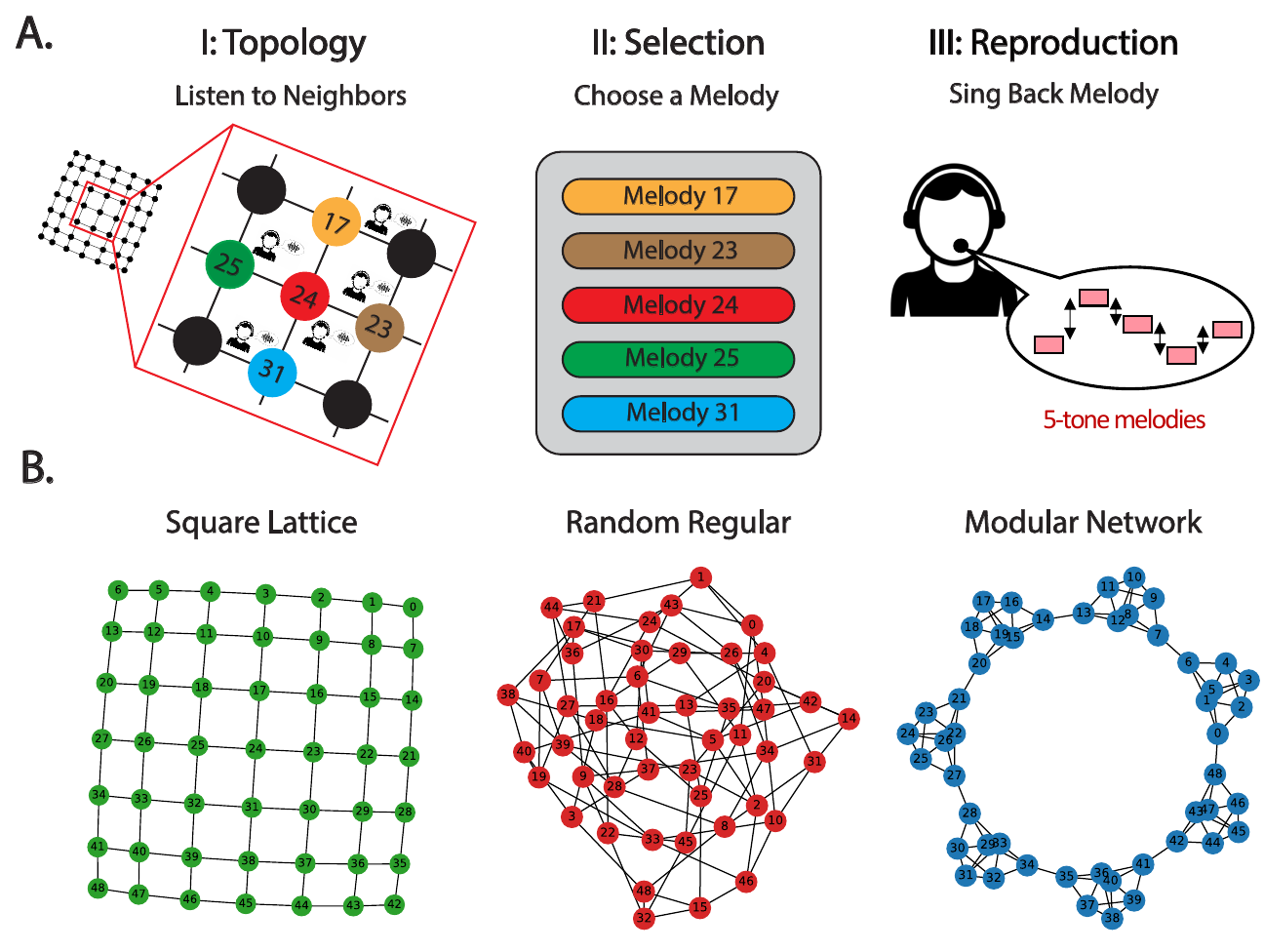}
\end{center}
\vspace{-2mm}
\caption{Schematic of the paradigm. \textbf{A.} The three components involved in the cultural process. \textbf{B.} The three topologies considered, a square lattice (periodic boundary conditions not depicted), a random regular graph, and a modular network.} 
\label{fig:schematic-mechanism}
\end{figure}

When examining the evolution of human cultural artifacts, three key mechanisms are often discussed: topology, selection, and reproduction (Figure \ref{fig:schematic-mechanism}A). Topology refers to the structure of the social network, determining the efficiency at which information spreads \cite{centola2022network}. Selection determines what information is more likely to be adopted by individuals, often due to social learning strategies, such as conformity bias, prestige, or aesthetic appeal \cite{kendal2018social}. Reproduction refers to the cognitive and physiological biases of learners, such as inductive biases, vocal constraints, and memory limits, which shape how individuals transform information as they pass it on \cite{xu2010rational,anglada2023large}. Although previous research has discussed how such mechanisms may affect the evolution of cultural systems, it has thus far been difficult to evaluate their relative importance and how their interaction shapes cultural evolution. This is in part due to a critical methodological limitation: achieving a balance between a naturalistic domain with strong biases, and a scalable experimental design that enables a controlled ablation of different factors is challenging.

Here, we address this challenge by integrating recent advances in online experimental methods \cite{HarrisonMarjieh2020,anglada2023large} with a highly naturalistic task: singing, a widespread form of musical expression across cultures \cite{mehr2019universality}. Participants are organized into social networks and asked to select and reproduce melodies from their neighbors by singing (Figure \ref{fig:schematic-mechanism}A). Their creations are then passed on to the next generations of participants over multiple iterations. Over time, musical structures emerge and evolve, allowing us to probe causal links between underlying mechanisms and the properties of the evolving artifacts. Crucially, this paradigm achieves a balance between an ecologically valid domain, and the ability to experimentally turn on and off underlying mechanisms in the social system in a scalable and controlled way.

We apply this paradigm to three equally-sized network topologies that are locally identical but globally different (Figure~\ref{fig:schematic-mechanism}B): a square lattice (with periodic boundaries), a random regular network, and a modular network with different cliques. To study the impact of different mechanisms, we compare the evolution on these networks against a linear setting without topology, as well as a series of ablation studies where selection and reproduction biases are eliminated through controlled randomization. To assess the quality of creations, we use computational metrics and large-scale validation studies with independent human raters. Our results show that introducing selection and topology yields more complex and pleasant repertoires relative to those found in the linear condition, as well as systematic variations across topologies. Moreover, these topological differences are severely diminished in the ablation studies, suggesting an interaction between mechanisms. These findings pave the way towards a fuller understanding of how culture is shaped by the conjunction of social and cognitive processes.

\begin{figure*}[ht]
\begin{center}
\includegraphics[width=0.8\linewidth]{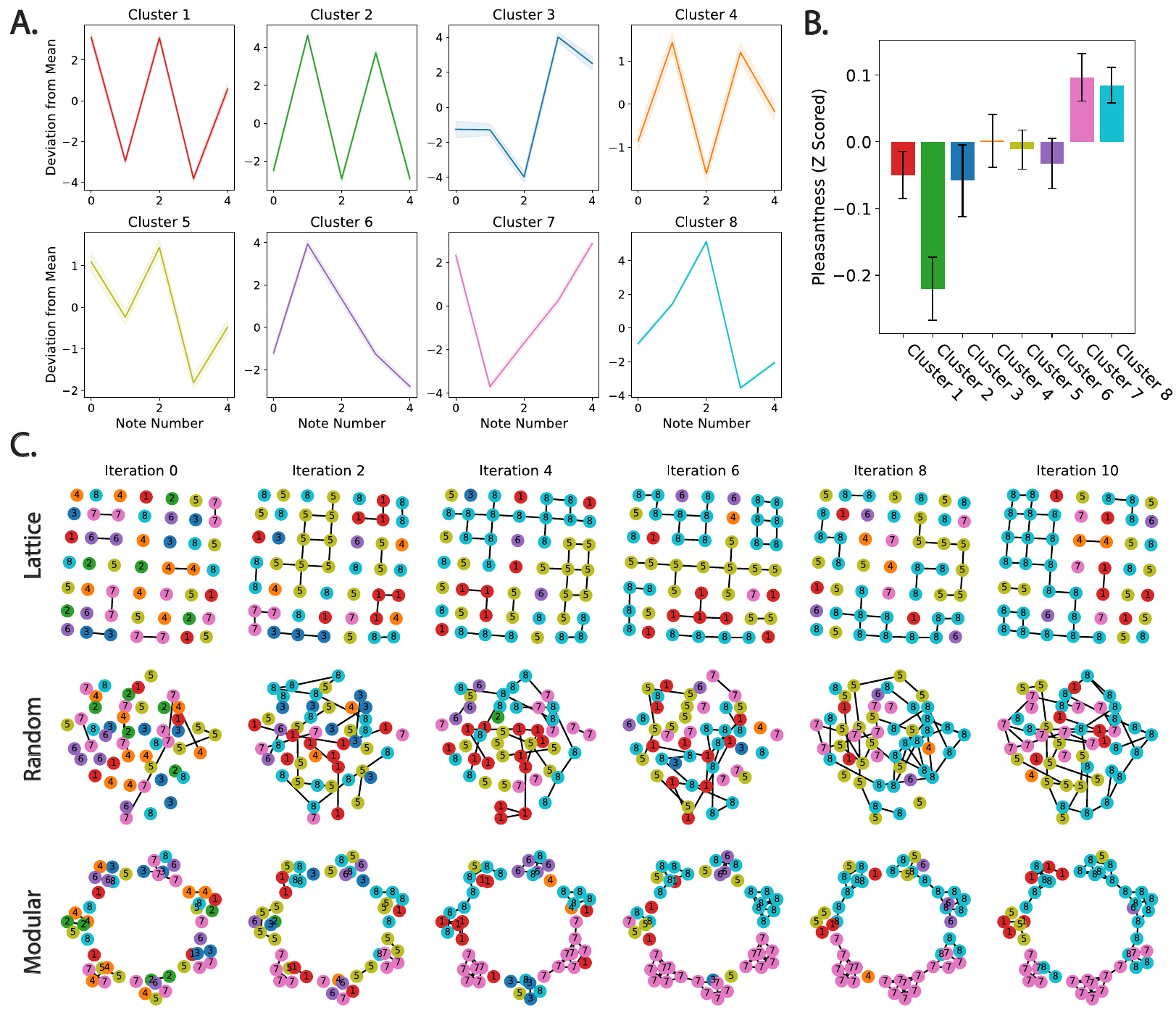}
\end{center}
\vspace{-4mm}
\caption{Emergent melodic prototypes. \textbf{A.} Melodic contours derived from jointly clustering all data (deviation from mean is given in semitones). \textbf{B.} Mean z-scored pleasantness ratings for each melodic cluster. \textbf{C.} Example evolution of prototypes as a function of time from one experimental batch. Nodes are colored based on the cluster of their melody, and edges are highlighted when the neighboring nodes share the same cluster (see Methods). Error bars indicate 95\% confidence intervals (CIs).} 
\label{fig:clusters}
\end{figure*}

\section{Background}
\subsection{Mechanisms of Cultural Evolution}

Recent advances in computational and experimental techniques have yielded insights into how individual mechanisms affect the evolution of cultural systems. First, iterated learning and serial reproduction experiments, where information is passed sequentially from one participant to another (akin to the `telephone game') show that, in the absence of selection and topology, reproduction biases essentially determine the asymptotic distribution of artifacts in a population \cite{kalish2007iterated,xu2010rational,kirby2014iterated,jacoby2024commonality}.  

Second, research on social learning strategies demonstrates that selection biases can depend on both content (e.g., a melody’s pleasantness) and context (e.g., how frequently the melody appears in the environment) \cite{kendal2018social}. Selection has also been shown to help preserve more complex artifacts that otherwise would not persist \cite{thompson2022complex}. However, there is ongoing debate about selection’s precise role, with two competing views with respect to its importance relative to reproduction \cite{mesoudi2021cultural}.

Finally, research suggests that a key network property that impacts global performance is ‘informational efficiency’, typically operationalized in terms of average minimum path length \cite{centola2022network} (i.e., the number of ‘steps’ required to traverse from one node to another in a network). For example, it has been shown that, counterintuitively, global performance on complex tasks is improved by reducing the network’s informational efficiency \cite{derex2016partial}. 

An important limitation of the prior literature is that the different mechanisms are rarely studied together, leaving a critical gap in our understanding of how their interactions may contribute to cultural evolution. Such interactions are important because cultural artifacts rarely evolve under the influence of a single mechanism. Moreover, previous studies have largely focused on artificial tasks, such as learning abstract languages, or solving problems in computer games. This limits our ability to generalize conclusions to naturalistic domains that are more complex, higher dimensional, and entail stronger biases that are shaped by real-life experience.

\subsection{Singing and Serial Reproduction}
Human song exhibits remarkable structural diversity while also sharing certain common features across cultures \cite{mehr2019universality}. 
Cross-cultural commonalities in human song have long been thought to derive from universal constraints on pitch perception, memory, and vocal production \cite{savage2015statistical}. However, much less is known about how features of the transmission process (e.g., vocal production) and the social environment (e.g., network structure) affect the evolution of music diversity and complexity. 
A recent study leveraged the paradigm of serial reproduction to conduct online singing experiments, where melodies were orally transmitted across hundreds of participants \cite{anglada2023large}. The results showed that reproduction biases led to the emergence of diverse musical structures, increasing the stability of music over time. However, as with other studies, the linear design makes it difficult to study the impact of selection and topology. Here, we overcome this challenge by adapting the singing paradigm into a social network.

\section{Singing Networks: Ablation Studies on a Cultural Process}
To characterize the interaction of topology, selection, and reproduction we introduce a behavioral paradigm whereby participants are asked to \emph{choose} and \emph{reproduce} five-note melodies from a location in a \emph{network} (Figure~\ref{fig:schematic-mechanism}A). Since the melodies available to each participant are themselves produced by other participants, iterating this process results in an evolving repertoire of melodies. By asking participants to choose and reproduce melodies, selection and reproduction biases are naturally incorporated, and by specifying the structure of the network (i.e., its topology) we can determine what information each location in the network has access to. 

Crucially, our design enables a controlled ablation of the various mechanisms. First, in the special case of a fully disconnected network, where melodies are transmitted sequentially from one participant to the next, our paradigm recovers the iterated singing paradigm of \citeA{anglada2023large}, providing a natural baseline to compare against (Study I). Second, by assigning participants random melodies from their environment to reproduce (instead of choosing themselves) we can eliminate the effect of selection while keeping reproduction and topology (Study II). Third, by reproducing selected melodies with artificial noise (instead of singing), we can eliminate the effect of reproduction while keeping selection and topology (Study III).
\begin{figure*}[t]
\begin{center}
\includegraphics[width=0.95\linewidth]{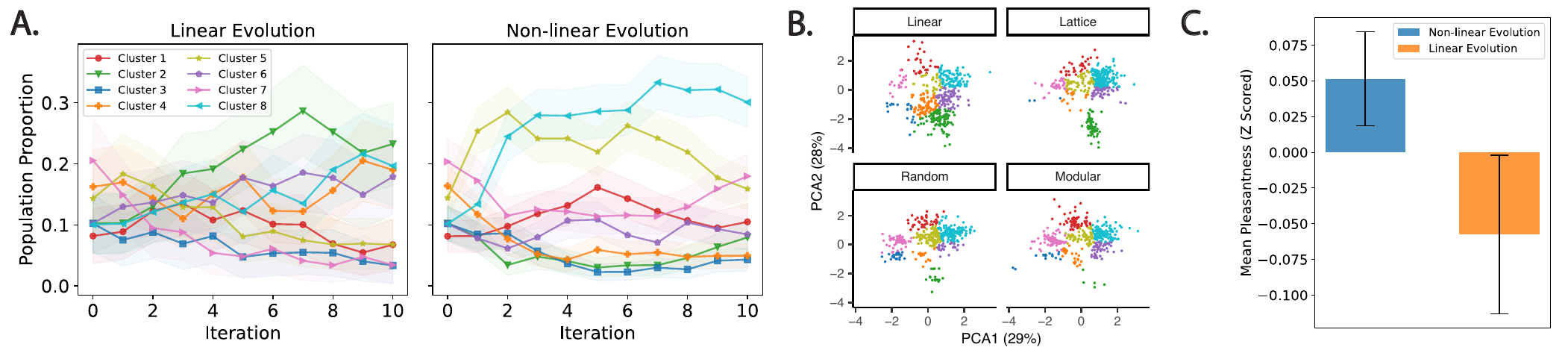}
\caption{Comparison between linear and non-linear (i.e., with topology) evolution. \textbf{A.} Cluster prevalence as a function of iterations. \textbf{B.} Clustered melodies in PCA space across last three iterations (see Methods). \textbf{C.} Average population pleasantness after a three-iteration burn in period (all conditions are initialized randomly and identically). Error bars indicate 95\% CIs.}
\label{fig:lin-vs-full}
\end{center}
\end{figure*}

For network topologies, we consider three types (Figure~\ref{fig:schematic-mechanism}B): a two-dimensional square lattice with periodic boundaries (i.e., a torus), a random regular network (i.e., with random edges and fixed degree), and a modular network, composed of several cliques on a ring (i.e., characterized by lower informational efficiency; see Methods). All networks had a fixed degree of four. These networks were chosen such that they are globally very different (see Methods), but also identical both in size and in terms of the local point of view of a participant since in each trial there are precisely five melodies to choose from (four from neighbors and one from the parent node). This enables a balanced comparison across topologies that eliminates any confounds that may arise from variation in task effort. Throughout, we accompany each experimental condition with a validation experiment in which the melodies are evaluated by an independent pool of raters. 

\section{Methods}
\subsection{Participants}
A total of 2,404 participants took part in the study ($N=1,931$ in the singing experiments; $N=473$ in the rating experiments). Participants ranged in age from 18 to 79 years ($M = 39.78$, $SD = 12.84$) and reported typical levels of musical training ($M = 2.46$, $SD = 1.35$), as measured by the Goldsmiths Musical Sophistication Index \cite{mullensiefen2013goldsmiths}. Recruitment was conducted online through Prolific\footnote{\url{www.prolific.com}}, targeting participants based in the UK.  
All participants provided informed consent in accordance with an approved protocol (Max Planck Ethics Council \#202142). Participants were compensated at a rate of £9 per hour. 

\subsection{Networked Experiments Infrastructure}
The network infrastructure was designed using PsyNet\footnote{\url{www.psynet.dev}}, a modern framework for experiment design and recruitment \cite{HarrisonMarjieh2020}. Upon specifying the set of network topologies within a given experimental batch, PsyNet asynchronously manages the assignment of participants to different locations within the networks. Once a participant is assigned to a given location, that location is locked until the participant completes their trial. The response is then processed and recorded, and it then becomes available for the next iteration of participants that get assigned to locations that have access to that graph node. In our design, there were three topologies per experimental batch, and each participant completed only three trials per topology in a random order. All topologies were initialized with the same random set of melodies (see below), and we replicated each condition three times. This across-participant design builds on prior literature \cite{anglada2023large} and ensures (i) scalability, (ii) that the same participants take part in all topologies, and (iii) that their contribution to a given topology is small.

\subsection{Topologies}
We specified and visualized the graph topologies using \texttt{networkx}, a Python package for graph computations \cite{hagberg2008exploring}. All three topologies had 49 nodes, and each node was connected to 4 neighbors (Figure~\ref{fig:schematic-mechanism}B). The graph edges were undirected meaning that information could propagate both ways. The square lattice was constructed using \texttt{grid\_2d\_graph} and had an average path length of $\bar{L}=3.5$, and betweenness-centrality of $\bar{c}=.053$. The random regular graph was constructed using \texttt{random\_regular\_graph} and unlike the grid its edges were connected at random subject to the constraint of fixed degree. Its graph metrics were $\bar{L}=2.84$ and $\bar{c}=.039$. Finally, the modular network \cite{lynn2020humans} was constructed by first creating 7 disjoint cliques of 4-rings, each comprising 7 nodes (i.e., 7 nodes are first organized on a ring and then connected to their 4 nearest neighbors on the ring), then one edge was removed per clique and replaced with two edges connecting to two other cliques while preserving the degree (Figure~\ref{fig:schematic-mechanism}B). The metrics in this case were $\bar{L}=5.92$ and $\bar{c}=.105$.

\subsection{Online Iterated Singing Paradigm}
We used the automated online pipeline for iterated signing experiments developed in \citeA{anglada2023large}. Melodies were parametrized as lists of numbers using MIDI notation, which maps frequency to a semitonal logarithmic scale (e.g.,  the middle C in a piano keyboard is mapped to the MIDI note number 60). To generate the initial melodies (shared across all experiments), we randomly sampled notes from a uniform pitch range of $[-15, 15]$ semitones. Consequently, initial melodies were completely random and did not follow any specific rules from music theory or scale systems. 

Participants' vocal productions were recorded in the browser and automatically analyzed, synthesized, and played to the next participant as the input melody, using a pitch-roving technique to minimize inter-trial dependencies and adjust melodies to participants’ singing range. The fundamental frequency ($f_0$) of notes in participants' vocal recordings was estimated using an automated method for pitch extraction based on \texttt{parselmouth} \cite{jadoul2018introducing} (see \citeA{anglada2023large} for details).
Melody tones in all experiments were 550 msec long and separated by 250 msec of silence. They were synthesized using a complex harmonic timbre and played using Tone.js\footnote{\url{https://tonejs.github.io/}}, a Web Audio framework for sound generation in the browser.

\subsection{Ablation Studies}
\noindent\textbf{Linear Baseline.} The first ablation study provided a way to compare our paradigm to the linear paradigm of \citeA{anglada2023large}. It was identical to the design of the networked (full) condition but with a trivial topology in which there are no edges between nodes and hence no transmissions between neighbors or selection.   

\noindent\textbf{No Selection Condition.} To eliminate the effect of selection we replicated the full condition but instead of asking participants to sing back the melody they selected, we provided them with an alternative melody that was randomly sampled from the available set in their local environment and asked them to sing it back.

\noindent\textbf{No Reproduction Condition.} To eliminate the effect of reproduction bias, we repeated the full condition but instead of propagating the sung melody we propagated the one that was selected with the addition of matched Gaussian noise. The matched Gaussian noise was generated as follows. First, we took all target melodies and their sung responses from the full condition, and computed the difference between their relative intervals, i.e., $\Delta_{st}=(i_{s1}-i_{t1},,i_{s2}-i_{t2},i_{s3}-i_{t3},i_{s4}-i_{t4})$ where $i_s$ and $i_t$ are the relative intervals of the sung and target melodies, respectively. Then, we computed the empirical covariance matrix associated with these four dimensions $\Sigma_\Delta$. We then used this matrix to add multivariate Gaussian noise $\mathcal{N}(0,\Sigma_\Delta)$ to the selected melodies in the ablation condition. This way we captured the general error scale and correlations but without any systematic deviations. 

\subsection{Procedure}
We used a combination of techniques to ensure high data quality \cite{anglada2022repp, anglada2023large,woods2017headphone}, including audio calibration and recording tests, headphone checks, and a singing performance test to measure participants' general singing abilities, and to exclude participants who could not provide minimal working data, as well as detecting participants' singing range (low vs.~high). In the singing experiments, participants were presented with five-tone melodies and asked to reproduce them by singing. Their reproductions were then synthesized on the fly to generate the stimuli for the next participants. Participants completed a total of 9 trials per experimental batch (3 trials per topology in networked experiments). Trials that did contain five detected notes were failed until a new participant could provide a valid response. Participants were not aware that they were interacting with other participants or that they were taking part in a social network experiment. In the rating experiments, participants were presented with 50 melodies in random order and asked to rate how `pleasant' they sounded, on a scale from 1 (`very unpleasant') to 7 (`very pleasant').

\subsection{Data Analysis}
\noindent\textbf{Clustering.}
We applied a clustering procedure to group melodies based on melodic prototypes, which capture recurring patterns of ups and downs in pitch that define a melody’s identity. First, we combined data from all singing experiments into a single dataset (15,726 melodies). To account for differences in absolute pitch, we subtracted the mean pitch of each melody from its notes, ensuring that clustering was based on relative pitch patterns. Next, we applied Principal Component Analysis (PCA) to the relative pitch values and retained the first two principal components, which together explained 57\% of the variance. We then performed k-means clustering \cite{macqueen1967some} to identify distinct melodic clusters. The optimal solution consisted of eight clusters, as determined by a silhouette score of .63 \cite{rousseeuw1987silhouettes}.

\noindent\textbf{Population Statistics.} We computed two population metrics associated with the clustering scheme above. The first is cluster entropy which measures population diversity. Let $p_i$ denote the fraction of cluster $i$ at a given iteration. Cluster entropy is then defined as $H(p)=-\sum_ip_i\log p_i$. The second metric is neighbor similarity which measures the fraction of neighbors that share the same cluster at a given iteration. Mathematically, we have $s=\sum_{(i,j)\in E}\delta_{c_i,c_j}/|E|$ where $\delta$ is an indicator that yields $1$ when the clusters of node $i$ and $j$ are equal and zero otherwise, and $E$ is the edge set. To measure systematic variations across topologies irrespective of global shifts that may arise from different participant groups across batches, we first computed the relative values of $H$ and $s$ with respect to their mean at each iteration (within a batch) and then aggregated across batches and iterations.

\begin{figure}[t]
\begin{center}
\includegraphics[width=0.9\linewidth]{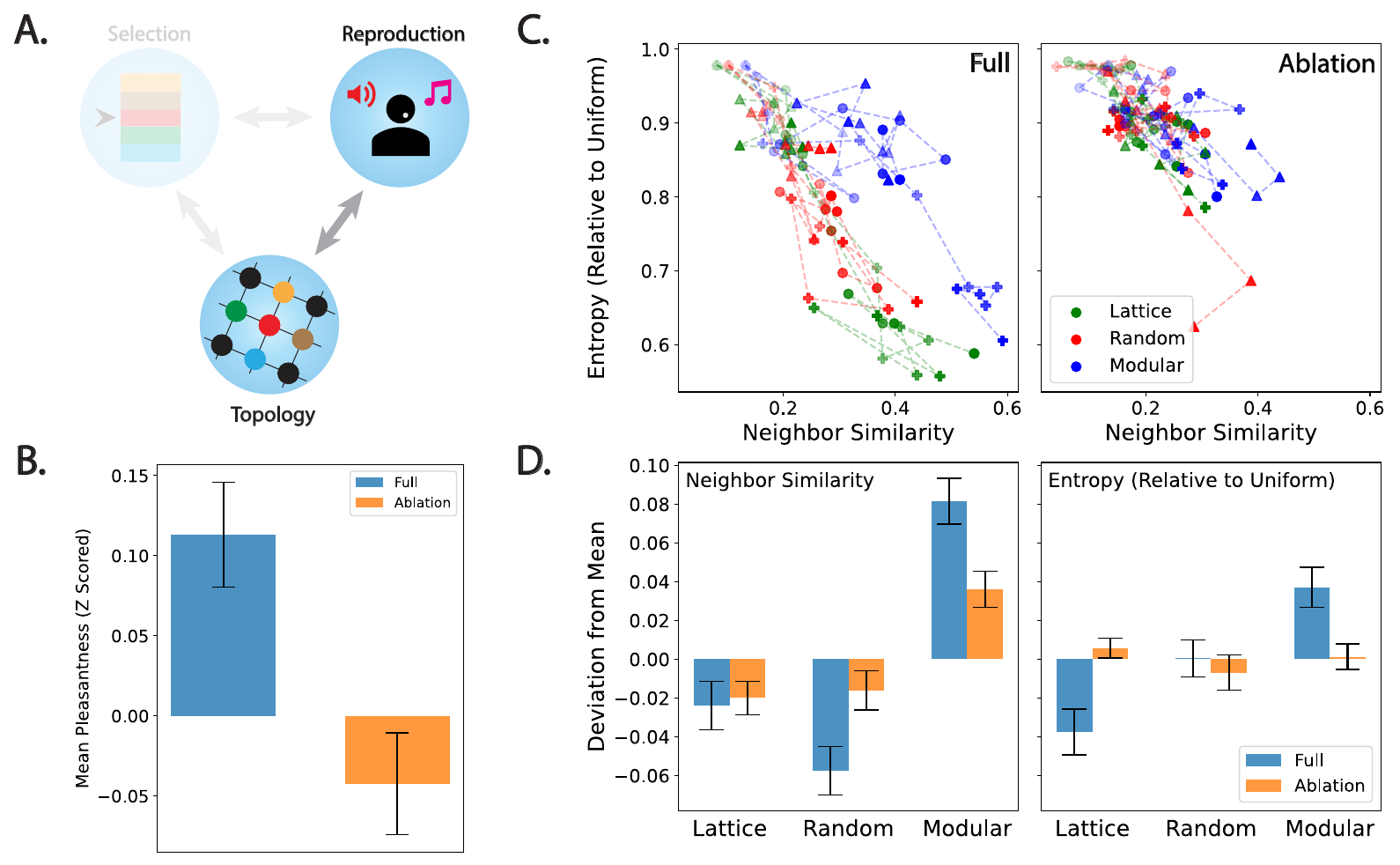}
\end{center}
\vspace{-4mm}
\caption{No selection condition. \textbf{A.} Melodies are sampled uniformly from the local environment for imitation. \textbf{B.} Average population pleasantness after a three iteration burn-in. \textbf{C.} Evolution of neighbor similarity and entropy. Markers indicate different experimental batches (per condition). More transparent colors indicate earlier iterations. \textbf{D.} Average deviation from mean for neighbor similarity and entropy after a three iteration burn-in. Error bars indicate 95\% CIs.}
\label{fig:ablation-selection}
\end{figure}

\section{Results}
\subsection{Study I: Non-linear versus Linear Evolution}
The emergent melodic prototypes and their average pleasantness ratings are shown in Figure~\ref{fig:clusters}A-B. We see that the contours differ both in their shape and in their perceived pleasantness. For example, cluster 2 exhibits a repeating upward-downward interval pattern ($+-+-$) but is perceived as the least pleasant (95\% CI $[-.27,-.17]$). On the other hand, clusters 7 and 8 are non-repeating ($-+++$ and $++-+$, respectively) but are perceived as the most pleasant (95\% CI: $[.06,.13]$, and $[.06,.11]$, respectively), possibly because they align with the patterns found in musical cadences. 

Figure~\ref{fig:clusters}C shows an example of the evolution of contours from one batch. We see that while initially the distribution of clusters is relatively uniform, by the last iteration some clusters take over (e.g., cluster 8). To make this precise, we computed the average cluster proportion (across all batches) in the full condition and compared that against the proportions found in the linear condition. The results are shown in Figure~\ref{fig:lin-vs-full}A. We see that while in the linear evolution case the proportions remained relatively intermingled by the last iteration (with cluster 2 being the most common; 95\% CIs at iterations 10 and 0 are $[.16,.30]$ and $[.05,.15]$, respectively; see also Figure~\ref{fig:lin-vs-full}B), the non-linear condition was able to support a stable growth of cluster 8 (95\% CIs at iterations 10 and 0 were $[.26,.34]$ and $[.07,.13]$, respectively). Note also that cluster 8 is significantly more pleasant than cluster 2 (pleasantness 95\% CIs are $[-.27,-.17]$ and $[.06,.11]$, respectively), which was also consistent with the significant difference in the general population pleasantness (Figure~\ref{fig:lin-vs-full}C).

\subsection{Study II: Turning off Selection}
Next we analyze the impact of selection by comparing the full condition against an ablation condition (Figure~\ref{fig:ablation-selection}A; see Methods) whereby melodies are selected at random from each neighborhood for reproduction (rather than being selected by the individual). Starting from pleasantness (Figure~\ref{fig:ablation-selection}B), we found that randomizing selection has a significant effect on the pleasantness of the produced melodies (pleasantness 95\% CIs are $[.08,.15]$ and $[-.07,-.01]$ for the full and ablation condition, respectively), suggesting that individuals do not merely choose melodies that are easy for reproduction but ones that are also pleasant.

In addition to population pleasantness, we considered neighbor similarity and entropy (see Methods). Neighbor similarity captures the extent to which neighboring nodes share the same melodic cluster (i.e., correlation). Entropy measures how diverse are the clusters in the population. Note that high neighbor similarity does not imply low entropy (e.g., when multiple melodic cliques emerge). The dynamics of the raw metrics are shown in Figure~\ref{fig:ablation-selection}C. We see that in the full condition different topologies occupy different parts of the space, with the modular network yielding generically higher neighbor similarity and entropy relative to the other topologies within the same batch. These differences, however, appear to be severely diminished in the ablation condition.

Quantitatively, the results are shown in Figure~\ref{fig:ablation-selection}D. We found that in the full condition the modular network yielded higher neighbor similarity than the other topologies (95\% CIs: $[.070,.093]$, $[-.070,-.045]$, $[-.036,-.011]$ for modular, random regular, and lattice, respectively) and that this effect was diminished when selection was removed (95\% CIs: $[.027,.045]$, $[-.026,-.010]$, $[-.029,-.011]$). A similar pattern was observed in the case of entropy whereby the modular network yielded the highest relative value, but this effect was eliminated in the ablation condition (compare full condition CIs: $[.027,.047]$, $[-.009,.010]$, $[-.049,-.026]$ for modular, random regular, and lattice, respectively, vs. ablation CIs: $[-.005,.001]$, $[-.016,.002]$, $[.001,.011]$). These findings suggest that selection supports the emergence of musical cliques in the modular network (as seen in Figure~\ref{fig:clusters}C). 

\subsection{Study III: Turning off Reproduction}
\begin{figure}[t]
\begin{center}
\includegraphics[width=0.9\linewidth]{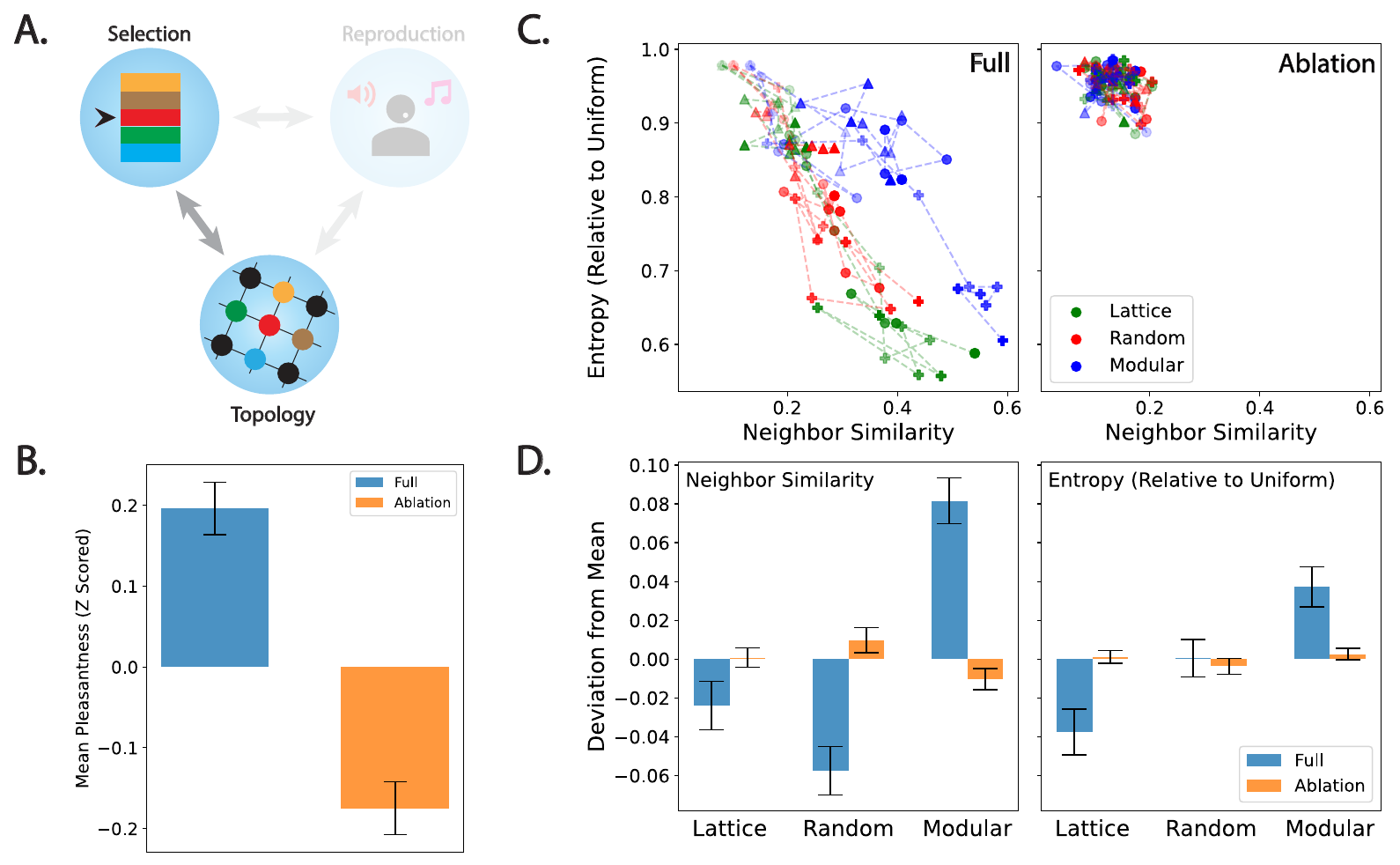}
\end{center}
\caption{No reproduction condition. \textbf{A.} Selected melodies are mutated through Gaussian noise rather than singing. \textbf{B.} Average population pleasantness. \textbf{C.} Evolution of neighbor similarity and entropy. \textbf{D.} Average deviation from mean for neighbor similarity and entropy. See Figure~\ref{fig:ablation-selection} for details.}
\label{fig:ablation-reproduction}
\end{figure}

Finally, we consider the ablation condition whereby instead of propagating selected melodies through singing we mutated them by adding Gaussian noise that lacks systematicity but is otherwise matched in terms of deviation size relative to the data in the full condition (see Methods). Figure~\ref{fig:ablation-reproduction}B shows the average population pleasantness for the two conditions. We see that even though participants were able to choose their melodies in the ablation condition, without an appropriate inductive bias the general quality of melodies is compromised (pleasantness 95\% CIs were $[.16,.23]$ and $[-.21,-.14]$ for the full and ablation conditions, respectively).

As for the other metrics, Figures~\ref{fig:ablation-reproduction}C-D reveal a drastic effect of the ablation intervention on the dynamics, collapsing the differences across topologies. Indeed, the deviation from the mean CIs for neighbor similarity in the ablation condition were $[-.016,-0.005]$, $[.003,.016]$, $[-.004,.006]$ for modular, random regular, and lattice topologies, respectively. Likewise, for entropy those were $[-.000,.006]$, $[.008,.000]$, and $[-.002,.004]$. In both cases, these deviations are much smaller than those found in the full condition (these values are identical to those provided in the previous section).

\section{Discussion}
Our results highlight a strong interdependency between the mechanisms underlying the evolution of cultural artifacts. First, by introducing topology and selection, we showed how these lead to the emergence of more pleasant and complex artifacts relative to the typically studied linear baseline. This is consistent with recent work suggesting that selective social learning preserves complexity \cite{thompson2022complex}, and extends it to a naturalistic and subjective domain. Second, we showed that variation in topology can impact population characteristics, with the modular network resulting in distinct musical cliques. These effects, however, were severely diminished when either selection or reproduction were removed, suggesting that they jointly contribute to the observed topological effects. This is consistent with theoretical work showing that when selection is random topological differences become secondary \cite{whalen2017adding}. It also informs the debate on the relative importance of selection and reproduction \cite{mesoudi2021cultural}, suggesting that both are necessary.

We end by discussing some limitations. First, we assumed that the structure of the social network is fixed, but there are important cases in which the structure evolves in time (e.g., social media networks). Second, we considered topologies that are locally identical to ensure a uniform participant experience, however, variable degree networks play key roles in social phenomena, with small-world networks being a noteworthy case \cite{watts1998collective}. Third, our paradigm avoids any direct communication between participants or the transmission of agent information, however, in reality this is not the case and it may have important implications. We hope to explore these avenues in future work. 
  
\paragraph{Acknowledgments}
This research project was made possible with the support of the NOMIS Foundation.

\bibliographystyle{apacite}

\setlength{\bibleftmargin}{.125in}
\setlength{\bibindent}{-\bibleftmargin}

\bibliography{main}

\end{document}